\documentclass[amsmath,amssymb,nofootinbib,10pt,notitlepage,twocolumn]{revtex4-1}

\usepackage[utf8]{inputenc}
\usepackage{graphicx}
\usepackage{amsmath}
\usepackage{mathtools}
\usepackage[a4paper, top=2.5cm, bottom=2.5cm, left=2.0cm,
right=2.0cm]{geometry}

\usepackage{hyperref}

\newcommand{\Tr}{\operatorname{Tr}}
\renewcommand{\Im}{\operatorname{Im}}

\begin{document}
    
\title{Effective elastic moduli of composites with a strongly disordered host material}
\date{\today}
\author{D.\,A.~Conyuh}
\author{A.\,A.~Semenov}
\author{Y.\,M.~Beltukov}
\affiliation{Ioffe Institute, Politechnicheskaya Str. 26, 194021 St. Petersburg, Russia}

\begin{abstract}
    The local elastic properties of strongly disordered host material are investigated using the theory of correlated random matrices. A significant increase in stiffness is shown in the interfacial region, which thickness depends on the strength of disorder. It is shown that this effect plays a crucial role in nanocomposites, in which interfacial regions are formed around each nanoparticle. The studied interfacial effect can significantly increase the influence of nanoparticles on the macroscopic stiffness of nanocomposites. The obtained thickness of the interfacial region is determined by the heterogeneity length scale and is of the same order as the length scale of the boson peak. 
\end{abstract}

\maketitle

\section{Introduction}

Amorphous glassy materials exhibit spatially inhomogeneous microscopic elastic properties due to their disordered structure~\cite{Yoshimoto-2004, Tsamados-2009, Wagner-2011, Mizuno-2013}. The local elastic heterogeneity results in non-affine deformations of amorphous solids under uniform stress. The presence of non-affine deformations was observed in a wide range of amorphous materials: metallic glasses~\cite{Jana-2019}, polymer hydrogels~\cite{Wen-2012}, supercooled liquids~\cite{Del-2008}, Lennard-Jones glasses~\cite{Goldenberg-2007}, and silica glass~\cite{Leonforte-2006}. The typical length scale of non-affine deformations was estimated as tens of particle sizes for Lennard-Jones glasses~\cite{Leonforte-2005}. For smaller length scales, the classical continuum elasticity theory can not be applied~\cite{Tanguy-2002}.

If the size of an amorphous medium is much larger than its heterogeneity length scale, one can use the macroscopic elastic moduli to describe the mechanical properties of this system. However, in composite systems containing amorphous materials, some regions may have small typical sizes. An important example is nanocomposites, in which the size of nanoinclusions may be comparable to the heterogeneity length scale of the host amorphous medium. Therefore, it is important to study the local elastic properties of amorphous solids, especially near the interface with other materials.

Amorphous polymers are an important class of amorphous materials. The elastic properties of polymer nanocomposites attract considerable interest due to their unique properties and great potential as future materials~\cite{Mai-2006, Thostenson-2003, Rafiee-2009, Mesbah-2009}. It was established that doping a polymer with nanoparticles, even at low concentrations, could lead to significant changes in the elasticity of the host material~\cite{Fu-2008, Ou-1998, Wang-2002, Wetzel-2003, Bershtein-2021}. 

It was proposed that the elastic properties of nanocomposites can be described by the so-called three-phase model~\cite{Odegard-2005}. The model assumes that the structure of a polymer is perturbed around the nanoparticle, which results in an effective interphase region around the nanoparticle with intermediate elastic properties. The interphase region has a strong influence on the macroscopic stiffness of the nanocomposite due to the large total surface area of nanoparticles. At present, the three-phase model is usually used as a phenomenologic model to fit the influence of inclusions on macroscopic elastic moduli obtained experimentally or using molecular dynamics~\cite{Bondioli-2005, Saber-2007, Qiao-2009, Wang-2011, Amraei-2019, Bazmara-2021}. 

Recent molecular dynamics studies have directly shown an increase in local elastic moduli of epoxy near the boehmite nanolayer~\cite{Fankhanel-2019} and polystyrene near the silica nanoinclusion~\cite{BeltukovPRE-2022}. In the latter case, an increase in polystyrene stiffness was revealed within a characteristic range of 1.4~nm from the nanoparticle, while polystyrene density saturates to the bulk value at significantly shorter distances. The enhancement of the local elastic properties of the polymer was attributed to the effect of non-affine deformations, which requires a more detailed theoretical study. 

It was shown that the general vibrational and mechanical properties of amorphous solids can be studied by the random matrix model~\cite{BeltukovPRB-2013, ConyuhFTT-2019}. Recently, using the theory of correlated random matrices, the analytical form of the vibrational density of states and the dynamical structure factor was obtained~\cite{ConyuhPRB-2021}. 

In the present paper, the theory of correlated random matrices is applied to study the effect of disorder on local elastic properties.

\section{Linear response}
\label{sec:response}

Macroscopic elastic properties determine the relationship between the macroscopic strain of a system and the applied macroscopic stress. In the general case, a linear response to some external force $f_i$ acting to $i$th degree of freedom of the system at frequency $\omega$ is determined by the following equation:
\begin{equation}
    \sum_j \bigl[\Phi_{ij} - \omega^2 m_{ij} \bigr] u_j = f_i,  \label{eq:lr-1}
\end{equation}
where $\smash{\hat{\Phi}}$ is the force-constant matrix, $\hat{m}$ is the mass matrix (usually $\hat{m}$ is a diagonal matrix, but we are not limited to this case), and $u_j$ is the displacement of $j$th degree of freedom from the equilibrium position. In the linear approximation, the force-constant matrix $\hat{\Phi}$ determines the linear elastic properties of a particular system. However, although the response $u_i$ is different for each realization of $\hat{\Phi}$, the fluctuations of macroscopic quantities based on $u_i$ (e.g. the macroscopic strain) are negligible. Therefore, it is important to find the average reaction $\langle u_i \rangle$, which can be expressed from Eq.~(\ref{eq:lr-1}) in the next form:
\begin{equation}
    \langle u_i \rangle = -\sum_{j} G_{ij} (\omega^2)f_j,   \label{eq:u_aver}
\end{equation}
where the resolvent
\begin{equation}
    \hat{G}(z) = \left< \frac{1}{\hat{m}z - \hat{\Phi}} \right> \label{eq:G}
\end{equation}
is introduced. The angle brackets denote the averaging over different realizations of $\hat{\Phi}$, and $z$ is a complex number. The relation between the average response $\langle u_j \rangle$ and the forces $f_i$ can be expressed as 
\begin{equation}
    \sum_j \bigl[\Phi^{\rm eff}_{ij}(\omega^2) - \omega^2 m_{ij} \bigr] \langle u_j \rangle = f_i,
\end{equation}
where $\hat{\Phi}^{\rm eff}$ is the effective force-constant matrix, which can be written using the resolvent $\hat{G}(z)$ as
\begin{equation}
    \hat{\Phi}^{\rm eff}(z) = \hat{m}z - \hat{G}(z)^{-1}.   \label{eq:Phi-eff}
\end{equation}
The study of the effective force-constant matrix $\hat{\Phi}^{\rm eff}$ is the main goal of the present paper. In such an analysis, the difference in each realization of $\hat{\Phi}$ must be taken into account. Note that for a strongly disordered system, the matrix $\hat{\Phi}^{\rm eff}$ may significantly differ from the mean force-constant matrix $\langle \hat{\Phi} \rangle$. To find the properties of $\hat{\Phi}^{\rm eff}$ we use the random matrix theory, which is based on the general properties of amorphous solids.

\section{Random matrix approach}

The force-constant matrix $\hat{\Phi}$ of an amorphous solid has some general properties. The most important one is the stability of the mechanical system, which means that the matrix $\hat{\Phi}$ is positive semi-definite. This condition is equivalent to the possibility to represent $\hat{\Phi}$ in the form
\begin{equation}
    \hat{\Phi} = \hat{A}\hat{A}^T,
\end{equation}
where $\hat{A}$ is some rectangular matrix \cite{Bhatia-2009}. The $i$th row of the matrix $\hat{A}$ corresponds to the $i$th degree of freedom. In the atomic system, each atom has three degrees of freedom. The $k$th column corresponds to the $k$th bond, which has the positive-definite quadratic potential energy \cite{BeltukovJETP-2016}
\begin{equation}
    U_k = \frac{1}{2}\left(\sum_i A_{ik} u_i\right)^2.
\end{equation}
In this paper, a system with $N$ degrees of freedom and $K$ bonds will be considered, which corresponds to $N\times K$ matrix $\hat{A}$. 

Each bond may involve several degrees of freedom, which affects the number of non-zero elements in $k$th column of the matrix $\hat{A}$. Therefore, the number and positions of non-zero elements in the matrix $\hat{A}$ depend on the type of interaction between atoms in an amorphous solid. For example, in the case of two-body potential (e.g. Lennard-Jones potential), each bond involves six degrees of freedom. In the case of three-body potential (e.g. Stillinger-Weber potential \cite{Stillinger-1985}), each term, which depends on the covalent bond angle, involves nine degrees of freedom. 

For an amorphous solid, the matrix $\hat{A}$ has a random nature. One can assume that matrix elements $A_{ij}$ are random numbers (some of them may be zero). However, due to the fact that the strongly disordered system is near the stable equilibrium on the verge of stability loss \cite{Alexander-1998}, there is a correlation between the elements of the matrix $\hat{A}$. In the general case, this correlation is described by the pair correlations $\langle A_{ik} A_{jl}\rangle = C_{ij, kl}$. Angle brackets denote the averaging over different realizations of $\hat{A}$. 

The effective force constant matrix $\hat{\Phi}^{\rm eff}$ is related to the correlation matrix $\hat{C}$. This relation is obtained in Appendix \ref{sec:aver} in the assumption that the elements of the matrix $\hat{A}$ are Gaussian random numbers with zero mean. The result given in Appendix \ref{sec:aver} is a generalization of the averaging method described in \cite{Burda-2004}.

Different bonds have different positions in space and involve different sets of degrees of freedom, which is described by the covariance matrix $\hat{C}$. Each column of the matrix $\hat{A}$ may have its own covariance matrix $\langle A_{ik} A_{jk}\rangle = C_{ij}^{(k)}$. In this paper, different columns of the matrix $\hat{A}$ representing different bonds are assumed to be uncorrelated with each other, which corresponds to the covariance matrix of the form
\begin{equation}
    \langle A_{ik} A_{jl}\rangle = C_{ij}^{(k)} \delta_{kl}.
\end{equation}
This assumption allows to describe the effective elastic properties of amorphous solids in the most simple form. In the previous paper \cite{ConyuhPRB-2021}, a stronger assumption $\langle A_{ik} A_{jl}\rangle = C_{ij} \delta_{kl}$ was applied, which did not take into account the difference between covariance matrices $\hat{C}^{(k)}$ and could not be used to describe local elastic properties. 

Using the results of Appendix \ref{sec:aver}, the effective force-constant matrix can be presented as
\begin{equation}
    \hat{\Phi}^{\rm eff}(\omega^2) = \sum_k \gamma_k(\omega^2) \hat{C}^{(k)},    \label{eq:Meff}
\end{equation}
where $\gamma_k(\omega^2)$ characterizes the frequency-dependent dimensionless stiffness of $k$th bond and can be found from the following system of nonlinear equations:
\begin{equation}
    \gamma_k(z) = 1 + \Tr \biggl[\gamma_k(z)\hat{C}^{(k)} \Big(\hat{m}z - \sum_l \gamma_l(z){\displaystyle \hat{C}^{(l)}}\Big)^{-1} \biggr].  \label{eq:Ek}
\end{equation}
where $z$ is some complex number. In a general case, Eq.~(\ref{eq:Ek}) can be solved numerically for any set of covariance matrices $\hat{C}^{(k)}$. However, in some cases Eq.~(\ref{eq:Ek}) can be simplified, which is considered in the next section.

\section{Effective elastic medium}

In this section an amorphous solid with homogeneous statistical properties is considered.  For a such medium, one can assume a homogeneous distribution of $K$ bonds over a system with $N$ degrees of freedom. In this case one can introduce a smooth function $\gamma({\bf r}, z)$ such that $\gamma_k(z) = \gamma({\bf r}_k, z)$, where ${\bf r}_k$ is a coordinate of $k$th bond. 

In the volume of a pure macroscopic amorphous solid, $\gamma({\bf r}, z)$ does not depend on the coordinate ${\bf r}$. However, the boundary effects may lead to non-homogeneous $\gamma({\bf r}, z)$ near the boundaries of amorphous solids. In Appendix~\ref{sec:hom}, the differential equation for $\gamma({\bf r}, z)$ is derived. In the static case ($z=\omega^2 = 0$), the equation for $\gamma({\bf r}) \equiv \gamma({\bf r}, 0)$ reads as
\begin{equation}
    (1 + \varkappa)\gamma({\bf r}) = \varkappa + \xi_b^2\Delta \ln \gamma({\bf r}),  \label{eq:E}
\end{equation}
where $\varkappa = K/N - 1$, $\xi_b$ is a typical bond size, and $\Delta$ denotes the Laplacian. 

For slowly varying $\gamma({\bf r})$ in the region near the point ${\bf r}$, the effective dynamical matrix has the form $\hat{\Phi}^{\rm eff} = \gamma({\bf r}) \langle \hat{\Phi} \rangle$, where $\langle\hat{\Phi}\rangle = \sum_k \hat{C}^{(k)}$ is the averaged force constant matrix. Therefore, $\gamma({\bf r})$ can be considered as a dimensionless elasticity since elastic moduli of the reference medium described by $\langle\hat{\Phi}\rangle$ are multiplied by $\gamma({\bf r})$.

Far from boundaries in an amorphous solid $\gamma({\bf r}) = \gamma_0 = \varkappa/(\varkappa + 1)$. If the number of random bonds is much greater than the number of degrees of freedom ($K\gg N$ and $\varkappa \gg 1$), self-averaging of random bonds takes place. It results in small fluctuations of the force constant matrix $\hat{\Phi}$. In this case $\hat{\Phi}^{\rm eff}$ is close to the average force constant matrix $\langle\hat{\Phi}\rangle$, and $\gamma_0\approx 1$.  The opposite case $\varkappa \ll 1$ corresponds to a strongly disordered solid with $\gamma_0 \ll 1$. Therefore, the effective medium is much softer than the reference medium described by the average force constant matrix $\langle\hat{\Phi}\rangle$. The role of disorder controlled by the parameter $\varkappa$ on the vibrational properties of the bulk amorphous solid was studied in~\cite{ConyuhPRB-2021}.

Equation (\ref{eq:E}) can be written as
\begin{equation}
    \alpha({\bf r}) = 1 + \xi^2\Delta \ln \alpha({\bf r}),  \label{eq:E2}
\end{equation}
where $\alpha({\bf r}) = \gamma({\bf r})/\gamma_0$ is the effective local elastic contrast, and $\xi=\xi_b/\sqrt{\varkappa}$ is the only dimensional parameter in the above equation.

The effective local elastic contrast $\alpha({\bf r})$ specifies effective local elastic moduli of the amorphous medium at coordinate ${\bf r}$: the local effective bulk modulus is ${\cal K}({\bf r}) = \alpha({\bf r}){\cal K}_0$ and the local effective shear modulus is $\mu({\bf r}) = \alpha({\bf r})\mu_0$, where ${\cal K}_0$ and $\mu_0$ are the corresponding elastic moduli of a pure macroscopic amorphous solid. Near the boundaries, the effective local elastic contrast $\alpha({\bf r})$ may differ from 1. The length scale of the boundary effects is described by  $\xi$. Since $\xi \sim \varkappa^{-1/2}$, it depends on the strength of disorder. Therefore, $\xi$ represents the heterogeneity length scale of the amorphous system. For strongly disordered medium $\xi\gg\xi_b$.

To obtain $\alpha({\bf r})$ in the whole amorphous solid, Eq.~(\ref{eq:E2}) should be accomplished with the boundary conditions. The most important case is the interface of an amorphous medium with a more rigid and ordered medium. Such a rigid and ordered medium can be considered as a medium with $\varkappa \gtrsim 1$. In this case, one can assume $\gamma({\bf r}) \sim 1$ on the boundaries. For strongly disordered medium ($\varkappa \ll 1$), this boundary condition means $\alpha({\bf r}) = \gamma({\bf r})/\gamma_0 \sim 1/\varkappa \gg 1$. Therefore, without the loss of precision, one can assume that $\alpha({\bf r}) = \infty$ on the boundaries to solve Eq.~(\ref{eq:E2}). Below, the two most important geometries of the boundary of an amorphous body are considered.

\subsection{Flat boundary}

\begin{figure}[t]
    \includegraphics[scale=0.75]{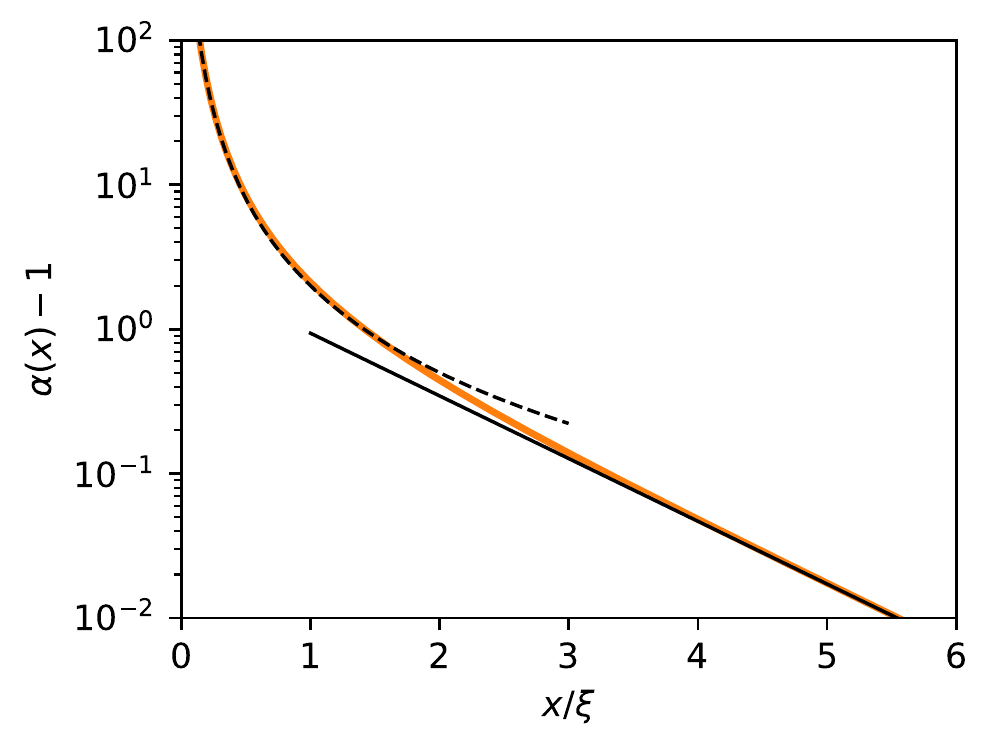}
    \caption{The effective local elastic contrast $\alpha(x)$ near the flat boundary as a function of the scaled distance to the boundary. Solid and dashed black lines show the asymptotics given by Eqs.~(\ref{eq:E_far_1D}) and (\ref{eq:E_close_1D}), respectively. } 
    \label{fig:1d}
\end{figure}

Near a flat boundary, $\alpha({\bf r})$ depends only on the distance from the boundary, which is denoted by $x$. In this case, Eq.~(\ref{eq:E2}) has the one-dimensional form
\begin{equation} 
    \alpha(x) = 1 + \xi^2 \frac{\partial^2 }{\partial x^2} \ln \alpha(x).   \label{eq:1d}
\end{equation}
The solution of Eq.~(\ref{eq:1d}) has a universal dependence on the scaled coordinate $x/\xi$, which is shown in Fig.~\ref{fig:1d}. Far away from the boundary ($x\gg \xi$), the asymptotic solution is
\begin{equation}
    \alpha(x) = 1 + c_1 e^{-x/\xi}, \label{eq:E_far_1D}
\end{equation}
where $c_1\approx 2.5527$. Near the boundary ($x \ll \xi$), the asymptotic solution is
\begin{equation}
    \alpha(x) = \frac{2 \xi^2}{x^2}.  \label{eq:E_close_1D}
\end{equation}
One can note that solution (\ref{eq:E_close_1D}) is inapplicable in the region $x \lesssim \xi_b$, where the assumption of the slow variation of $\alpha(x)$ on the length scale $\xi_b$ is violated. Thus, the actual near-boundary value of $\alpha(x)$ is $\alpha(\xi_b) \sim 1/\varkappa$.

\subsection{Spherical inclusion}

\begin{figure}[t]
    \includegraphics[scale=0.75]{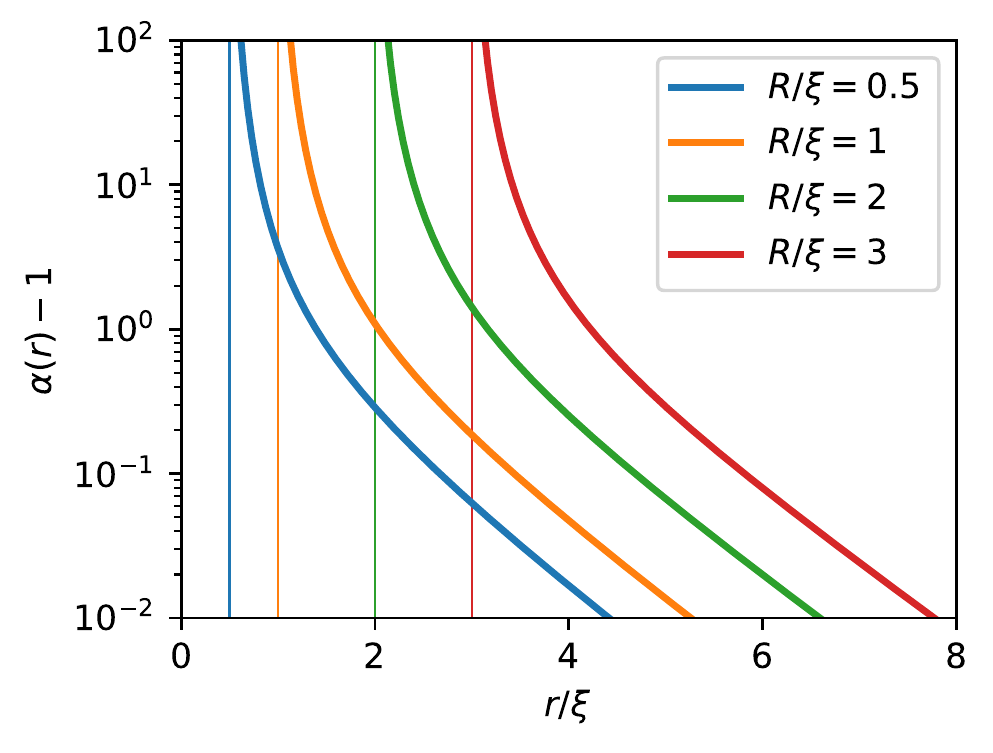}
    \caption{The effective local elastic contrast $\alpha(r)$ around the spherical nanoinclusion as a function of the scaled distance from the center of the nanoparticle for different nanoparticle radii. Thin vertical lines mark the corresponding radius of the nanoparticle.} 
    \label{fig:sphr}
\end{figure}

Another important example is spherical nanoinclusions in an amorphous medium. Around each nanoinclusions, Eq. (\ref{eq:E2}) can be written in spherical coordinates
\begin{equation} 
    \alpha(r) = 1 + \frac{\xi^2}{r^2} \frac{\partial }{\partial r} \left(r^2 \frac{\partial }{\partial r}\ln \alpha(r)\right),  \label{eq:sphr}
\end{equation}
where $r$ is the distance from the center of the nanoinclusion of the radius $R$. The solution of Eq.~(\ref{eq:sphr}) is shown in Fig.~\ref{fig:sphr}. Far away from the nanoinclusion ($r - R \gg \xi $), the asymptotic solution is
\begin{equation}
    \alpha(r) = 1 + c_2\frac{R}{r} e^{-(r-R)/\xi}, \label{eq:E_close_far}
\end{equation}
where $c_2$ is a coefficient, which depends on the ratio $R/\xi$. The asymptotic solution near the surface of nanoinclusion (${r - R \ll \xi, R}$) is
\begin{equation}
    \alpha(r) = \frac{2R^2 \xi^2}{r^2(r-R)^2}.   \label{eq:E_close_sphr}
\end{equation}
As in the one-dimensional case, solution (\ref{eq:E_close_sphr}) is inapplicable in a thin near-boundary region ${r - R \lesssim \xi_b}$. Thus, the actual near-boundary value of $\alpha(r)$ is $\alpha(R + \xi_b) \sim 1/\varkappa$.

Equations (\ref{eq:E_close_far}), (\ref{eq:E_close_sphr}) show that the effective elastic shell is formed around the spherical nanoparticle. The typical thickness of this shell is about the heterogeneity length scale $\xi$. Thus, the presence of the nanoinclusion enhances the elastic properties at a distance $\xi$ from the nanoparticle.

\section{Elastic properties of nanocomposite}
\label{sec:num}

The macroscopic elastic properties of nanocomposite describe a response (a strain) to macroscopic stress applied to the nanocomposite. For a nanocomposite with an amorphous host material, local strains exhibit large fluctuations. However, the macroscopic strain has negligible fluctuations. Therefore, as was shown in Section \ref{sec:response}, the macroscopic strain will be the same if the amorphous material will be substituted with the effective elastic medium. Thus, the macroscopic elastic properties of nanocomposite with amorphous host material can be found in two steps: (i) find the non-random continuous effective medium described by $\alpha({\bf r})$ and (ii) find the macroscopic elastic properties of the nanocomposite with effective medium using the classical elasticity theory.

In this Section, we demonstrate this approach for nanocomposite with rigid spherical inclusions in the host amorphous matrix. For simplicity of the calculation, the inclusions are placed in sites of a simple cubic lattice with period $L$. In this case, Eq.~(\ref{eq:E2}) can be solved in one periodic cubic cell $L \times L \times L$ with one rigid spherical inclusion of radius $R$ placed in the center of the cell.

In this Section, the finite element method with the hexagonal mesh containing $N_1 = 37888$ elements is used, which was described in detail in~\cite{SemenovJSS-2020}. FEniCS v0.5.2~\cite{Alnaes-2015} is used to solve the finite element problem using variational formulation.

\begin{figure*}
	\includegraphics[scale=0.75]{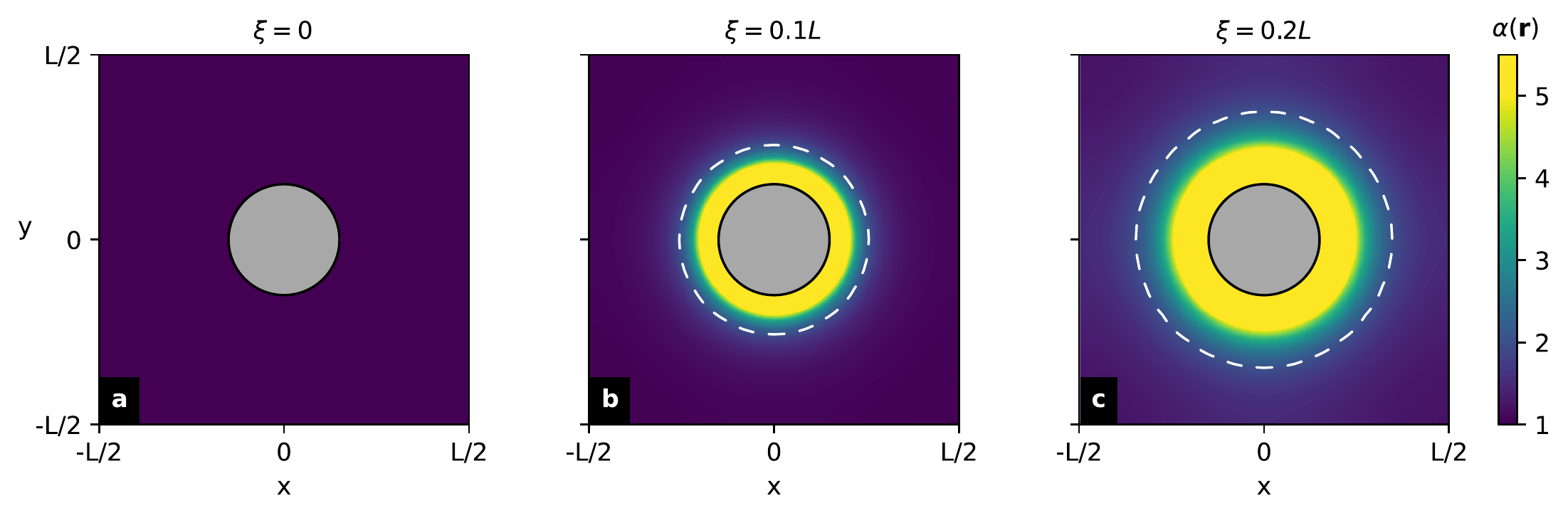}
	\caption{The distribution of the effective local elastic contrast $\alpha({\bf r})$ for samples with different values of the heterogeneity length scale $\xi$. The section passing through the center of a spherical rigid inclusion of radius $R=0.15L$ (indicated by a gray circle) is shown. The white dotted line shows the shell on which the effective contrast $\alpha({\bf r})=2$. The yellow color shows the area where the effective contrast $\alpha({\bf r}) > 5$.} 
	\label{fig:halo}
\end{figure*}

In step (i), the effective local elastic contrast $\alpha({\bf r})$ is found using Eq.~(\ref{eq:E2}) on the mesh under consideration. Figure \ref{fig:halo} shows the obtained spatial distribution of effective local elastic contrast $\alpha({\bf r})$ for different heterogeneity length scale $\xi$ in the plane passing through the center of the inclusion of radius $R=0.15 L$.

In step (ii), the classical elasticity theory is used to find the macroscopic elastic properties of the nanocomposite. Such macroscopic properties determine the relationship between the macroscopic strain tensor of nanocomposite $\varepsilon^{\rm nc}_{ij}$ and the macroscopic stress tensor of nanocomposite $\sigma^{\rm nc}_{ij}$:
\begin{equation}
	\sigma^{\rm nc}_{ij} = {\cal C}_{ijkl}^{\rm nc} \varepsilon^{\rm nc}_{kl}. \label{eq:Hooke}
\end{equation}
The macroscopic strain tensors $\varepsilon^{\rm nc}_{ij}$ and stress tensors $\sigma^{\rm nc}_{ij}$ are a simple averaging over the composite volume of the corresponding local tensors:
\begin{gather}
	\varepsilon^{\rm nc}_{ij} = \overline{\varepsilon_{ij}(\mathbf{r})},\\
	\sigma^{\rm nc}_{ij} = \overline{\sigma_{ij}(\mathbf{r})}.
\end{gather}

The relation between the local strain and stress tensors is determined by local elastic moduli ${\cal K}({\bf r}) = \alpha({\bf r}){\cal K}_0$ and $\mu({\bf r}) = \alpha({\bf r})\mu_0$:
\begin{multline}
	\sigma_{ij}(\mathbf{r}) = {\cal K}({\bf r})\delta_{ij} \varepsilon_{kk}(\mathbf{r}) \\
    + 2\mu({\bf r})\biggl(\varepsilon_{ij}(\mathbf{r}) - \frac{1}{3}\delta_{ij} \varepsilon_{kk}(\mathbf{r})\biggr) .
	\label{eq:sigma local}
\end{multline}
To determine the macroscopic elasticity tensor ${\cal C}_{ijkl}^{\rm nc}$, it is necessary to determine such stresses and strains that satisfy the force balance equation
\begin{equation}
	\frac{\partial}{\partial r_i} \sigma_{ij} (\bold{r}) = 0,
	\label{eq:sigma eq}
\end{equation}
and boundary conditions.
For a composite with periodically spaced inclusions, the boundary conditions can be satisfied by considering the displacement in the form
\begin{equation}
	u_i(\bold{r}) = \varepsilon^{\rm nc}_{ij}r_j+\tilde{u}_i(\bold{r}),
	\label{eq: gran eq1}
\end{equation}
where $\tilde{u}_i(\bold{r})$ is a periodic function
\begin{multline}
	\tilde{u}_i(x,y,z) = \tilde{u}_i(x+L,y,z) = \\ \tilde{u}_i(x,y+L,z) = \tilde{u}_i(x,y,z+L).
	\label{eq: gran eq2}
\end{multline}
Thus, for a given macroscopic strain $\varepsilon^{\rm nc}_{ij}$, we can solve the elasticity equations (\ref{eq:sigma local}), (\ref{eq:sigma eq}) with boundary conditions (\ref {eq: gran eq1}), (\ref{eq: gran eq2}) and find the macroscopic stress $\sigma^{\rm nc}_{ij}$. Using a set of different macroscopic deformations $\varepsilon^{\rm nc}_{ij}$, all components of the macroscopic elasticity tensor $C_{ijkl}^{\rm nc}$ can be determined.

Elasticity equations were solved numerically for samples with different values of inclusion volume fraction $\phi = \frac{4}{3}\pi R^3/L^3$ and the length scale $\xi$.  For a small volume fraction of inclusions, the composite may be considered isotropic with a bulk modulus ${\cal K}_{\rm nc}$ and the shear modulus $\mu_{\rm nc}$:
\begin{equation}
    {\cal C}_{ijkl}^{\rm nc} = {\cal K}_{\rm nc} \delta_{ij} \delta_{kl}
    + \mu_{\rm nc} \left(\delta_{ik}\delta_{jl}+\delta_{il}\delta_{jk}- \frac{2}{3} \delta_{ij}\delta_{kl}\right).
\end{equation}
For a large volume fraction of inclusions, one should take into account the cubic anisotropy of the composite due to the periodic placement of inclusions. However, this anisotropy is not important for the effect under consideration and, therefore, is out of the scope of this paper.

\begin{figure}
	\includegraphics[scale=0.75]{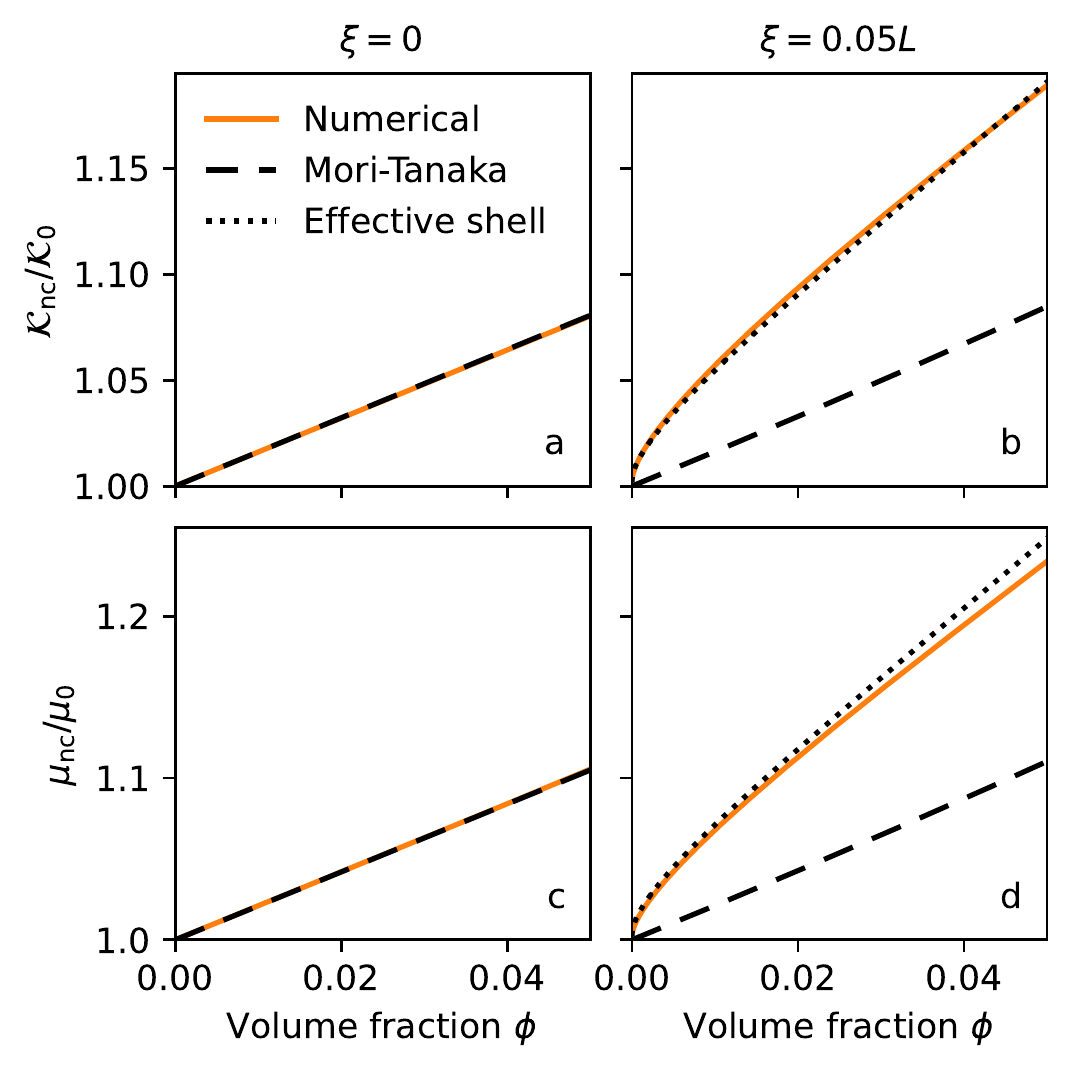}
	\caption{The ratio of the nanocomposite elastic moduli to the corresponding moduli of the host amorphous medium  depending on the volume fraction of rigid spherical inclusions. The volume fraction of inclusions is varied by the nanoinclusion radius with the fixed placement of nanoinclusions. Solid lines are the result of numerical simulation, dashed lines are the Mori-Tanaka model for rigid inclusions with radius $R$, and dotted lines are the modified Mori-Tanaka model for rigid inclusions with effective inclusion radius $R^{\rm eff} = R+1.2\xi$.}
	\label{fig:Ciis}
\end{figure}

Figure \ref{fig:Ciis} shows the results of calculating the reinforcement of amorphous medium due to spherical rigid inclusions. For an amorphous matrix, Poisson's ratio was chosen as $\nu_0 = 0.3$, which is a typical value for amorphous polymers. 


For homogeneous host material without disorder ($\xi=0$), the macroscopic stiffness of the nanocomposite can be calculated using the Mori-Tanaka approach~\cite{Mori-1973, Benveniste-1987}. The addition of a small concentration of rigid spherical inclusions to the host material leads to the following macroscopic elastic moduli of the nanocomposite:
\begin{align}
    {\cal K}_{\rm MT} &= {\cal K}_0\left(1 + 3 \phi \frac{1 - \nu_0}{1 + \nu_0}\right), \\
    \mu^{}_{\rm MT} &= \mu^{}_0\left(1 + \frac{15\phi}{2} \frac{1 - \nu_0}{4 - 5\nu_0}\right).
\end{align}
Figure \ref{fig:Ciis}a,c shows the perfect match between the numerical calculation and the Mori-Tanaka theory for the case $\xi=0$.

For amorphous host material (see Fig.~\ref{fig:Ciis}b,d for $\xi=0.05L$), the macroscopic elastic moduli ${\cal K}_{\rm nc}$ and $\mu_{\rm nc}$ of nanocomposite are significantly larger than the prediction by the Mori-Tanaka theory. Figure \ref{fig:halo} shows that around each nanoparticle there is an effective shell with enhanced elastic properties. The thickness of this shell is approximately the heterogeneity length scale $\xi$. Therefore, we plot the additional dotted lines in Fig.~\ref{fig:Ciis}b,d with the Mori-Tanaka theory but with increased nanoparticle radius $R^{\rm eff} = R+1.2\xi$. One can see a good agreement with a such modification of the existing theory. Factor 1.2 was chosen for better fitting of the result.

\section{Discussion}

The obtained effective force constant matrix $\hat{\Phi}^{\rm eff}$ can be used as a non-random substitution of the random force constant matrix $\hat{\Phi}$, which gives the same average response to external forces. Particularly, $\hat{\Phi}^{\rm eff}$ can be used to represent the macroscopic elastic properties of composite materials containing amorphous materials. 

The same effective force constant matrix $\hat{\Phi}^{\rm eff}$ can be used to find the \emph{average} response to microscopical forces at \emph{any} length scale. At the same time, one can expect strong fluctuations of atomic displacements for length scales below $\xi$. The analysis of fluctuations is the subject of future work.

Using the random matrix theory, the effective force constant matrix $\hat{\Phi}^{\rm eff}$ was obtained as a sum of covariance matrices $\hat{C}^{(k)}$ with some coefficients defined by Eq.~(\ref{eq:Ek}). In an amorphous system, bonds usually have a finite range, which results in sparse matrices $\hat{C}^{(k)}$. The results of the random matrix are applicable if the number of nonzero elements is much bigger than one. This assumption works reasonably well since the interaction between atoms involves two or more atoms, each of which has three degrees of freedom. Additionally, the numerical random matrix model described in~\cite{ConyuhPRB-2021} was analyzed and compared with the present theory. The effective force constant matrix obtained numerically quickly converges to the present theoretical predictions with the increase of the radius of bonds.

Equation (\ref{eq:Meff}) shows that the effective force-constant matrix $\hat{\Phi}^{\rm eff}$ describes a short-range interaction if matrices $\hat{C}^{(k)}$ describes finite-range bonds. This property holds for any strength of disorder, so the effective medium described by $\hat{\Phi}^{\rm eff}$ can be analyzed using the continuum elasticity theory.

For amorphous solids with homogeneous and isotropic statistics, Eq.~(\ref{eq:E2}) defines the static local stiffness of the effective medium. Near the boundary with a more rigid medium, the static local stiffness of the effective medium exceeds its bulk values. The heterogeneity length scale $\xi \sim \varkappa^{-1/2}$ defines the thickness of the boundary layer with increased stiffness. 

The studied effect is especially important for nanocomposites with an amorphous host material. In this case, one can find the macroscopic elastic moduli in two steps: (i) find the effective local elastic contrast $\alpha({\bf r})$ and (ii) use classical elasticity theory to find macroscopic elastic properties of the nanocomposite with the effective continuous medium. An example of this approach was presented in Section \ref{sec:num}.  The disorder of the host material leads to the formation of the effective shell of the thickness $\xi$ with increased stiffness around each nanoparticle. In this case, the nanoparticles have the effective radius $R_{\rm eff}$ such that $R_{\rm eff} - R \sim \xi$. Thus, for $R\sim\xi$ the influence of nanoinclusions on the macroscopic stiffness of the nanocomposite will be increased by an order of magnitude.

It is important that the studied effect is determined by the strength of the disorder. This effect should be distinguished from the effect of adhesion, which can directly change the density and other structural properties of the amorphous medium near the surface of nanoinclusions.

The main result (\ref{eq:E2}) concerns the static stiffness, however, Eqs. (\ref{eq:Ek}) and (\ref{eq:Erz}) may be applied to arbitrary frequency $\omega$ given by the parameter $z=\omega^2$. For amorphous solids with homogeneous and isotropic statistics, far away from boundaries $\gamma_k(z)$ does not depend on $k$ and can be written as $\gamma_k(z)=z/Z(z)$, where $Z(z)$ is some complex function. In this case $\hat{\Phi}^{\rm eff}(z) = \frac{z}{Z(z)}\sum_k \hat{C}^{(k)} = \frac{z}{Z(z)} \langle\hat{\Phi}\rangle$. Therefore, summing Eq.~(\ref{eq:Ek}) over $k$, we obtain the complex equation
\begin{equation}
    \varkappa Z(z) + \frac{Z(z)^2}{N} \Tr\left[ \frac{1}{Z(z) - \langle\hat{M}\rangle}\right] = (1+\varkappa)z,
\end{equation}
where $\langle\hat{M}\rangle = \hat{m}^{-1/2} \langle\hat{\Phi}\rangle \hat{m}^{-1/2}$ is the average dynamical matrix. For any given $z = \omega^2-i0$ one can find $Z(z)$ and obtain the vibrational density of states $g(\omega) = (1+\varkappa)\frac{2\pi}{\omega} \Im[1/Z(\omega^2-i0)]$. A more detailed analysis of vibrational properties has been done in~\cite{ConyuhPRB-2021}.

Amorphous solids have an excess of low-frequency vibrational density of states, known as the boson peak~\cite{Malinovsky-1986,Shintani-2008}. The boson peak length scale defined as $\xi_{\rm bp} = 2\pi c_T/\omega_{\rm bp}$, where $c_T$ is the transverse sound velocity and $\omega_{\rm bp}$ is the boson peak frequency, was attributed to the heterogeneity length scale~\cite{Leonforte-2005}. In the random matrix model, the boson peak length scale is $\xi_{\rm bp} \sim a_0 \varkappa^{-1/2}$, where $a_0$ is the interatomic distance~\cite{ConyuhPRB-2021}. Thus, the heterogeneity length scale $\xi$ and $\xi_{\rm bp}$ have the same order and the same dependence on the strength of disorder in the studied random matrix model.




In real amorphous solids, the strength of disorder cannot be varied in a wide range. However, model granular systems, known as jammed solids, have the possibility to change their properties significantly~\cite{OHern-2003}. This is due to a critical behavior of elastic and vibrational properties for small positive values of the parameter $z - z_0$~\cite{Wyart-2010, DeGiuli-2014, DeGiuli-2015}, which corresponds to the parameter $\varkappa$ in the present theory~\cite{ConyuhPRB-2021}. In jammed solids, the length scale $l_c \sim (z - z_c)^{-1/2}$ is related to the breakdown of the continuum elasticity~\cite{Lerner-2014} and coincides with the boson peak length scale~\cite{Wyart-2010}. Thus, the length scale $l_c$ corresponds to the length scale $\xi$ in the present theory. The investigation of the local elastic properties near the boundaries of jammed solids is of great interest to check the validity of Eq.~(\ref{eq:E2}) for such systems.

The obtained results are not limited to the study of the elastic properties of strongly disordered systems. Other properties requiring positive definiteness can be considered. For example, instead of stiffness, one can consider the conductivity of a strongly disordered medium. Thus, $\gamma({\bf r})$ can describe the increase of the conductivity of the effective medium near the interface with a well-conducting material. However, the applicability of the considered model to such kind of systems requires further research.

\section{Conclusion}

In this paper, the theory of the correlated random matrices was applied to find the local elastic properties of amorphous solids. The effective force constant matrix $\hat{\Phi}^{\rm eff}(z)$ was obtained, which can be used to find the average linear response to a force of a given frequency $\omega$ given by the parameter $z=\omega^2$.

For amorphous solids with homogeneous and isotropic statistical properties, a continuous  equation for effective local elastic contrast $\alpha({\bf r})$ was obtained. It reveals the increase of the stiffness of amorphous solid near the boundary with a more rigid and ordered body. The typical thickness of the boundary layer with increased stiffness is $\xi \sim \varkappa^{-1/2}$. Far away from the boundaries $\alpha({\bf r})$ has an exponential decay to 1 with a typical length~$\xi$.

For the strongly disordered amorphous solids ${\varkappa \ll 1}$, the heterogeneity length scale $\xi$ is much larger than the typical interatomic size in the system. The scaling of $\xi$ with the strength of disorder emphasizes the role of disorder in the formation of the boundary layer with increased stiffness.

The effect under study is important for macroscopic elastic moduli of nanocomposites with the amorphous host material. The numerical model of an amorphous solid with rigid spherical inclusions was studied to demonstrate the effect. It was shown that the shell with enhanced elastic properties is formed around each nanoparticle. The thickness of this shell is of the order of $\xi$, which results in the increased effective radius of nanoparticles, which significantly increases the macroscopic elastic moduli of the nanocomposite.

\section{Acknowledgments}

The financial support from Russian Science Foundation under the grant \#22-72-10083 is gratefully acknowledged. Authors thank A.V. Shumilin for the valuable discussions.

\onecolumngrid
\appendix

\section{Random matrix theory: the averaging procedure}
\label{sec:aver}

\newcommand{\diag}[2][0pt]{\mathord{\raisebox{#1}{\includegraphics[page=#2,scale=0.22]{diagrams6.pdf}}}}
\newcommand{\eqmathbox}[2][T]{\eqmakebox[#1]{$\displaystyle#2$}}

The averaging in the resolvent $G(z) = \big<(\hat{m}z - \hat{\Phi})^{-1}\big>$ can be done analytically for $\hat{\Phi} = \hat{A}\hat{A}^T$ where $\hat{A}$ is a Gaussian random matrix. In the general case, the matrix elements are correlated: $\langle A_{ik} A_{jl}\rangle = C_{ij, kl}$. The resolvent $G(z)$ can be presented as an infinite series
\begin{equation}
    \hat{G}(z) = \left<\frac{1}{\hat{m}z - \hat{A}\hat{A}^T}\right> 
    = \frac{1}{\hat{m}z} + \left< \frac{1}{\hat{m}z}\hat{A}\hat{A}^T\frac{1}{\hat{m}z} \right> + \left< \frac{1}{\hat{m}z}\hat{A}\hat{A}^T\frac{1}{\hat{m}z}\hat{A}\hat{A}^T\frac{1}{\hat{m}z} \right> + \cdots
    \label{eq:G-series}
\end{equation}
The elements of the resolvent $\hat{G}(z)$ can be written explicitly in the next form:
\begin{multline}
    G_{ij}(z) = (\hat{m}z)^{-1}_{ij} + \sum_{i_1k_1i_2k_2} (\hat{m}z)^{-1}_{ii_1} \delta^{\vphantom{-1}}_{k_1k_2} (\hat{m}z)^{-1}_{i_2j} \left< A_{i_1k_1}A_{i_2k_2} \right> + \\
    \sum_{i_1k_1i_2k_2i_3k_3i_4k_4} (\hat{m}z)^{-1}_{ii_1} \delta^{\vphantom{-1}}_{k_1k_2} (\hat{m}z)^{-1}_{i_2i_3} \delta^{\vphantom{-1}}_{k_3k_4} (\hat{m}z)^{-1}_{i_4j} \left< A_{i_1k_1}A_{i_2k_2}A_{i_3k_3}A_{i_4k_4} \right> + \cdots
    \label{eq:G-indices}
\end{multline}
We follow from the diagram technique described in \cite{Burda-2004} and introduce the next graphical representation:
\begin{align*}
    (\hat{m}z)^{-1}_{ij} = \diag[-7pt]{1}, \quad \delta_{kl} = \diag[-7pt]{2}, \quad \langle A_{ik} A_{jl}\rangle = C_{ij, kl} = \diag[-7pt]{3}.
\end{align*}
Here the solid line joining $i$ and $j$ is the factor $(\hat{m}z)^{-1}_{ij}$, the dashed line joining $k$ and $l$ is the Kronecker symbol $\delta_{kl}$, and a double arc joining $i,k$ and $l,j$ is the propagator $C_{ij, kl}$. Following these rules, the second term in (\ref{eq:G-indices}) corresponds to the next diagram:
\begin{align*}
    \sum_{i_1k_1i_2k_2} (\hat{m}z)^{-1}_{ii_1} \delta^{\vphantom{-1}}_{k_1k_2} (\hat{m}z)^{-1}_{i_2j} \left< A_{i_1k_1}A_{i_2k_2} \right> = \diag[-7pt]{4}.
\end{align*}
Since the elements of the matrix $\hat{A}$ are Gaussian random numbers, Wick's probability theorem is applicable for consecutively calculating even-point correlation functions, which are expressed as sums of all distinct products of two-point functions $\langle A_{i_1k_1} A_{i_2k_2}\rangle$:
\begin{align*}
    \langle A_{i_1k_1}A_{i_2k_2}A_{i_3k_3}A_{i_4k_4} \rangle 
    &= \langle A_{i_1k_1}A_{i_2k_2}\rangle \langle A_{i_3k_3}A_{i_4k_4} \rangle + \langle A_{i_1k_1}A_{i_4k_4}\rangle \langle A_{i_2k_2}A_{i_3k_3} \rangle + \langle A_{i_1k_1}A_{i_3k_3}\rangle \langle A_{i_2k_2}A_{i_4k_4} \rangle\\
    &= \diag[-7pt]{5} + \diag[-7pt]{6} + \diag[-7pt]{7}.
\end{align*}
Therefore, a graphical representation of the resolvent $\hat{G}(z)$ is
\begin{equation}
    \diag{8} = 
    \diag{9} + \diag{10} + \diag{11} + \diag{12} + \diag{13} + \dots.    \label{eq:G-diagram}
\end{equation}

The presentation \eqref{eq:G-diagram} allows us to distinguish planar and non-planar diagrams. For planar diagrams, the number of closed loops (closed solid line or closed dashed line) is equal to the number of double arcs. For non-planar diagrams, the number of closed loops is less than the number of double arcs. Namely, the second diagram in \eqref{eq:G-diagram} is planar and contains one closed loop and one double arc, the third and fourth diagrams are planar and contain two closed loops and two double arcs, and the fifth diagram is non-planar and contains two double arcs and only one closed loop. 

Each closed loop $\mathcal{L}$ corresponds to the calculation of a trace, which gives some factor $T_\mathcal{L}$ depending on the number of non-zero elements of the matrix $\hat{A}$. If each bond involves a sufficiently large number of degrees of freedom (although the matrix $\hat{A}$ can be a highly sparse matrix), the factor $T_\mathcal{L} \gg 1$ for each closed loop $\mathcal{L}$. In the case of a sufficiently filled matrix $\hat{A}$, the factor $T_\mathcal{L} \sim N$. At the condition $T_\mathcal{L} \gg 1$, each planar diagram contributes much more than a non-planar diagram with the same number of double arcs. Therefore, we can exclude non-planar diagrams from the summation \eqref{eq:G-diagram} and take into account only planar diagrams.

One can draw $\hat{G}(z)$ using the self-energy $\hat{\Sigma}(z)$ which contains only planar diagrams:
\begin{equation}
    \diag{8} = \diag{9} + \diag{14} + \diag{15} + \dots .    \label{eq:G-sum}
\end{equation}
The matrix $\hat{\Sigma}(z)$ can be expressed using the other resolvent $\hat{G}^\star(z) = \big<\big(1 - \hat{A}^T(\hat{m}z)^{-1}\hat{A}\big)^{-1}\big>$ by the Dyson–Schwinger relation:
\begin{equation}
    \diag{16} = \diag{17}.   \label{eq:Sum}
\end{equation}
The resolvent $\hat{G}^\star(z)$ contains all diagrams of the same shape as in (\ref{eq:G-diagram}) with dashed and solid lines replaced. Therefore, analogically to Eq.~(\ref{eq:G-sum}), it can be written as
\begin{equation}
    \diag{18} = \diag{19} + \diag{20} + \diag{21} + \dots,   \label{eq:G*-sum}
\end{equation}
where the self-energy $\hat{\Sigma}^\star(z)$ is related to $\hat{G}(z)$ by the Dyson–Schwinger relation:
\begin{equation}
    \diag{22} = \diag{23}.   \label{eq:Sum*}
\end{equation}
As a result, we obtain the closed set of four equations that correspond to the graphical representation (\ref{eq:G-sum})-(\ref{eq:Sum*}): 
\begin{alignat}{2}
    \hat{G}(z) &= \frac{1}{\hat{m}z - \hat{\Sigma}(z)}, \qquad
    & \Sigma_{ij}(z) &= \sum_{kl} C_{ij,kl} G_{kl}^\star(z), \label{eq:G-closed1} \\ 
    \hat{G}^\star(z) &= \frac{1}{\displaystyle 1 - \hat{\Sigma}^\star(z)},
    & \Sigma^\star_{kl}(z) &= \sum_{ij} C_{ij,kl} G_{ij}(z).
    \label{eq:G-closed2}
\end{alignat}

As it follows from equation (\ref{eq:Phi-eff}), the matrix $\hat{\Sigma}(z)$ plays the role of an effective force-constant matrix $\hat{\Phi}^{\rm eff}(z)$ describing the properties of an effective medium: 
\begin{equation}
    \Phi^{\rm eff}_{ij}(z) = \Sigma_{ij}(z).
\end{equation}
For any given covariance matrix $C_{ij,kl}$, one can solve Eqs.~(\ref{eq:G-closed1})--(\ref{eq:G-closed2}) and find the effective force-constant matrix $\hat{\Phi}^{\rm eff}(z)$.

In the case of uncorrelated bonds the matrix elements $A_{ik}$ and $A_{jl}$ are independent for $k \neq l$. The corresponding covariance matrix is
\begin{equation}
    C_{ij,kl} = C_{ij}^{(k)} \delta_{kl}.
\end{equation}
In this case, $\hat{G}^\star(z)$ has a diagonal form, $G^\star_{kl}(z) = \gamma_k(z)\delta_{kl}$, and the solution of Eqs.~(\ref{eq:G-closed1})--(\ref{eq:G-closed2}) can be presented in the following simplified form:
\begin{gather}
    \Phi^{\rm eff}_{ij}(z) = \sum_{k} C^{(k)}_{ij}\gamma_k(z), \\
    \gamma_k(z) = 1 + \Tr \biggl[\gamma_k(z)\hat{C}^{(k)} \Big(\hat{m}z - \sum_l \gamma_l(z){\displaystyle \hat{C}^{(l)}}\Big)^{-1} \biggr].
\end{gather}



\section{Amorphous solid with homogeneous and isotropic statistical properties}
\label{sec:hom}

One can assume that $\gamma_l(z) = \gamma({\bf r}_l, z)$ is close to $\gamma_k(z) = \gamma({\bf r}_k, z)$ for neighbor bonds $k$ and $l$. In this case, Eq. (\ref{eq:Ek}) can be written as
\begin{equation}
    \gamma_k(z) = 1 + W_k(Z) + \sum_l W_{kl}(Z) \frac{\gamma_l(z) - \gamma_k(z)}{\gamma_k(z)} 
    + \sum_{lm} W_{klm}(Z) \frac{(\gamma_l(z) - \gamma_k(z))(\gamma_m(z) - \gamma_k(z))}{\gamma_k^2(z)},
\end{equation}
where $Z = z/\gamma_k(z)$ and
\begin{align}
    W_k(Z) &= \Tr \biggl[\hat{C}^{(k)} \frac{1}{\hat{m}Z - \langle \hat{\Phi} \rangle } \biggr], \\
    W_{kl}(Z) &= \Tr \biggl[\hat{C}^{(k)} \frac{1}{\hat{m}Z - \langle \hat{\Phi} \rangle } \hat{C}^{(l)} \frac{1}{\hat{m}Z - \langle \hat{\Phi} \rangle } \biggr],\! \\
    \!W_{klm}(Z) &= \Tr \biggl[\hat{C}^{(k)} \frac{1}{\hat{m}Z - \langle \hat{\Phi} \rangle } \hat{C}^{(l)} \frac{1}{\hat{m}Z - \langle \hat{\Phi} \rangle } \hat{C}^{(m)} \frac{1}{\hat{m}Z - \langle \hat{\Phi} \rangle } \biggr].\!
\end{align}
At the same time, $\gamma_l(z) - \gamma_k(z)$ can be written as
\begin{equation}
    \gamma_l(z) - \gamma_k(z) = \sum_\alpha \frac{\partial \gamma({\bf r}_k, z)}{\partial r_\alpha} (r_{l\alpha} - r_{k\alpha}) 
    + \frac{1}{2}\sum_{\alpha\beta} \frac{\partial^2 \gamma({\bf r}_k, z)}{\partial r_\alpha \partial r_\beta} (r_{l\alpha} - r_{k\alpha})(r_{l\beta} - r_{k\beta}).
\end{equation}
As a result, the following differential equation for $\gamma({\bf r}, z)$ is obtained:
\begin{align}
    \gamma({\bf r}, z) = 1 + W({\bf r}, Z)
    &+ \frac{1}{\gamma({\bf r}, z)}\sum_\alpha W'_{\alpha}({\bf r}, Z)\frac{\partial \gamma({\bf r}, z)}{\partial r_\alpha}
    + \frac{1}{\gamma({\bf r}, z)}\sum_{\alpha\beta} W'_{\alpha\beta}({\bf r}, Z)\frac{\partial^2 \gamma({\bf r}, z)}{\partial r_\alpha \partial r_\beta} \\
    &+ \frac{1}{\gamma({\bf r}, z)^2}\sum_{\alpha\beta} W''_{\alpha\beta}({\bf r}, Z)\frac{\partial \gamma({\bf r}, z)}{\partial r_\alpha}\frac{\partial \gamma({\bf r}, z)}{\partial r_\beta},  \label{eq:Erz}
\end{align}
where
\begin{align}
    W({\bf r}_k, Z) &= W_k(Z), \\
    W'_\alpha({\bf r}_k, Z) &= \sum_l W_{kl}(Z)(r_{l\alpha} - r_{k\alpha}), \\
    W'_{\alpha\beta}({\bf r}_k, Z) &= \frac{1}{2}\sum_l W_{kl}(Z)(r_{l\alpha} - r_{k\alpha})(r_{l\beta} - r_{k\beta}), \\
    W''_{\alpha\beta}({\bf r}_k, Z) &= \sum_{lm} W_{klm}(Z)(r_{l\alpha} - r_{k\alpha})(r_{m\beta} - r_{k\beta}).
\end{align}
Static properties are defined by the limit $z\to 0$ and $Z \to 0$. In this case there are the following sum rules:
\begin{align}
    \sum_k W_k(0) &= N_0-N,  \label{eq:sumrule1} \\
    \sum_l W_{kl}(0) &= -W_k(0),  \label{eq:sumrule2} \\
    \sum_m W_{klm}(0) &= -W_{kl}(0),  \label{eq:sumrule3} 
\end{align}
where $N_0$ is the number of trivial zero-frequency modes (translations and rotations), which can be neglected for $N\gg 1$. Using Eqs. (\ref{eq:sumrule1})--(\ref{eq:sumrule3}), for an amorphous solid with homogeneous and isotropic statistical properties, one obtain
\begin{align}
    W({\bf r}_k, 0) &= -\frac{N}{K}, \\
    W'_\alpha({\bf r}_k, 0) &=  0,\\
    W'_{\alpha\beta}({\bf r}_k, 0) &= \frac{N}{K}\xi_b^2 \delta_{\alpha \beta}, \\
    W''_{\alpha\beta}({\bf r}_k, 0) &= -\frac{N}{K}\xi_b^2 \delta_{\alpha \beta},
\end{align}
where $\xi_b$ is a typical radius of the bonds. As a result, in the static case ($z=0$) we obtain
\begin{equation}
    \gamma({\bf r}, 0) = 1 - \frac{N}{K} + \frac{N}{K}\xi_b^2\Delta \ln \gamma({\bf r}, 0).
\end{equation}

\twocolumngrid
\bibliographystyle{apsrev4-1}
\bibliography{refs.bib}

\end{document}